\documentclass[twoside,a4paper,11pt]{article}

\usepackage{subfig,xurl}

\usepackage{irrj2e}

\newcommand{\added}[1]{#1}
\newcommand{\replaced}[2]{#1}

\usepackage{lastpage}

\irrjheading{1}{2025}{93--\pageref{LastPage}}{10/24; Revised 02/25}{05/25}{10.54195/irrj.19784}

\ShortHeadings{Supporting Evidence-Based Medicine by Finding Relevant and Significant Works}{Frihat and Fuhr}
\firstpageno{93}

\begin{document}

\title{Supporting Evidence-Based Medicine by Finding Both Relevant and Significant Works}

\author{\name Sameh Frihat \email sameh.frihat@uni-due.de \\
       \addr 
       University of Duisburg-Essen\\
       Duisburg, Germany
       \AND
       \name Norbert Fuhr \email norbert.fuhr@uni-due.de \\
       \addr 
       University of Duisburg-Essen\\
       Duisburg, Germany}

\editor{Haiming Liu}

\maketitle

\begin{abstract}
In this paper, we present a new approach to improving the relevance and reliability of medical information retrieval, which builds upon the concept of Level of Evidence (LoE)\@. The LoE framework categorizes medical publications into seven distinct levels based on the underlying empirical evidence. Despite LoE framework's relevance in medical research and evidence-based practice, only few medical publications explicitly state their LoE\@. Therefore, we develop a classification model for automatically assigning LoE to medical publications, which successfully classifies over 26 million documents in MEDLINE database into LoE classes. The subsequent retrieval experiments on the TREC Precision Medicine datasets show substantial improvements in retrieval relevance, when LoE is used as a search filter.
\end{abstract}

\begin{keywords}
Medical Document Facade, Level of Evidence, Evidence-Based Medicine, Medical Search Engines
\end{keywords}

\section{Introduction}

In medical research and practice, where findings and decisions directly impact human lives, successful retrieval of relevant and reliable information from scientific literature is paramount. Relevant information includes findings that are directly applicable to a condition under study, whereas reliable means that the findings are consistent under similar conditions~\citep{strage2023interobserver}. These concepts contribute to identifying significant information, which implies that findings have a practical and meaningful impact that is not due to chance in terms of its effect on patient care or outcomes~\citep{sathian2010relevance}. 

Modern evidence-based medicine (EBM) relies on a systematic approach to guide medical decisions using scientific evidence~\citep{burns2011levels, patrick2004evidence}. A key component of EBM is the Level of Evidence (LoE) framework, which categorizes medical research papers into 7 main distinct levels based on the strength and reliability of evidence reported~\citep{rosner2012evidence,desai2019hierarchy,van2023application}. This stratification, exemplified by the OCEBM (Oxford Centre for Evidence-Based Medicine {\small \url{https://www.cebm.net/}}) framework~\citep{howick2011oxford}, ranges from highly rigorous and reliable systematic reviews of randomized controlled trials (Level 1a) to case studies with limited evidential value (Level 4)~\citep{borawski2007levels,evidence2002users}.  

\noindent
Within this framework, each level holds unique significance, representing a specific study design and methodology \citep{borawski2007levels}. The hierarchy includes the following Levels of Evidence (LoEs):

\begin{itemize}
    \item \textbf{Level 1a: Systematic Reviews of Randomized Controlled Trials (RCTs)}.
At the apex of the LoE pyramid are systematic reviews and meta-analyses of well-conducted RCTs. Renowned for their comprehensive analysis of rigorous research, these reviews yield the most authoritative evidence.

    \item \textbf{Level 1b: Individual Randomized Controlled Trials (RCTs)}.
This level features individual RCTs that contribute crucial insights into causal relationships by evaluating interventions in controlled settings.

    \item \textbf{Level 2a: Systematic Reviews of Cohort Studies}.
Systematic reviews of cohort studies provide valuable evidence regarding associations between interventions and outcomes in real-world settings.

    \item \textbf{Level 2b: Individual Cohort Studies}.
Individual cohort studies at this level offer meaningful evidence about interventions' effects within specific populations.

    \item \textbf{Level 3a: Systematic Reviews of Case-Control Studies}.
Systematic reviews of case-control studies extend insight into the associations between interventions and outcomes, offering a broader perspective.

    \item \textbf{Level 3b: Individual Case-Control Studies}.
Individual case-control studies contribute evidence by exploring the relationships between interventions and outcomes within well-defined contexts.

    \item \textbf{Level 4: Case Series}.
At this level, case series provide preliminary evidence about interventions' effects, although they are limited by their susceptibility to biases and confounding factors.
\end{itemize}

\noindent
Although LoE is a crucial parameter for assessing a medical publication's significance, it is often not explicitly stated in publications, creating a problem for medical information retrieval (IR), where the aim is to retrieve significant medical publications or their content.

\added{
Our work addresses the `Acquiring' stage of the 5A's model (Ask, Acquire, Appraise, Apply, and Assess) in EBM~\citep{leung2001evidence}, which focuses on retrieving relevant literature to help users find the best available evidence. While LoE and the 5A's model are distinct frameworks, enabling users to filter retrieved information based on LoE supports the 'Acquiring' stage. Future work could explore integrating automated evidence appraisal to complement our retrieval approach.
}

In this article, we propose an automatic approach to identifying and prioritizing significant works in medical research. First, we develop a classification method for automatically assigning LoE to medical publications, then we use the identified LoE as a search filter in an IR setting. We demonstrate on the TREC PM (Precision Medicine) 2017--2019 collections~\citep{roberts2017overview} that using LoE as a filter when retrieving medical papers leads to improved retrieval results, and that the gain is highest for highly evidential medical papers.

\section{Related Work} \label{sec:relatedWork}
Recent advancements in Evidence-Based Medicine (EBM) have emphasized the role of automation in enhancing the classification and credibility assessment of Clinical Trials and RCTs. A key development in this area is the RobotReviewer system introduced by~\citep{marshall2014automating,marshall2016robotreviewer}, which automates the risk of bias assessment in RCTs and provides quality supporting text for bias assessments. This is vital for individual RCTs and also applicable to systematic reviews and meta-analyses of RCTs. The evaluation results indicate that RobotReviewer could match the performance of human reviewers in assessing the risk of bias~\citep{marshall2016robotreviewer,marshall2019toward}, which has been confirmed by several subsequent studies~\citep{soboczenski2019machine, hirt2021agreement, arno2022accuracy}. 
Further, \replaced{contributions from}{contributions from Hartling and Gates (2022)}~\citet{hartling2022friend} highlight the potential of such automation technologies to refine the quality and efficiency of systematic reviews, particularly in evaluating RCTs.

These advancements mark a significant shift in EBM, offering effective solutions for processing and categorizing extensive medical literature. However, these studies do not cover the full range of evidence levels of medical publications. Instead, they focus only on RCTs and their systematic reviews (Levels 1b and 1a in the LoE framework) and are possibly also applicable to levels 2b and 2a (cohort studies and their systematic reviews). 

\added{While large-scale metadata sources such as PubMed's ``publication type'' field offer broader coverage (e.g., labeling studies as ``Clinical Trial'' or ``Review''), they lack explicit evidence-hierarchy distinctions (e.g., differentiating high-quality RCTs from lower-quality observational studies) required for direct alignment with the LoE framework~\citep{pasche2020sib}. Several machine learning-based tools are developed and used for predicting the ``publication type'' field such as Anne O'Tate, RCT Tagger, Multi-Tagger, etc \citep{cohen2021fifty}. }

No other automation effort to date has explicitly attempted to incorporate the LoE framework, despite its central place in EBM practice. This work's main contribution is in providing a fully automatic retrieval system for medical publications by automatizing the EBM practice of assigning LoE to medical publications and then using LoE to decide on the relevance of a publication in a given context. 

\section{LoE Classifier} \label{sec:loeClassifier}
We view the problem of assigning LoE to medical publications as a classification task and explain in this section the training and the evaluation of the LoE classifier. 

\subsection{Data}

We use a dataset derived from the Oncology Guidelines of the German Association of Scientific Medical Societies (Arbeitsgemeinschaft der Wissenschaftlichen Medizinischen Fachgesellschaften\footnote{\url{https://www.awmf.org/}}). This dataset is unique in that it explicitly mentions the LoE of various medical publications as per the OCEBM framework. It includes 2816 publication--LoE pairs, extracted from unstructured PDFs\footnote{A structured format of the dataset is available on \url{https://github.com/samehfrihat/LevelOfEvidence}}. \added{The distribution of LoE levels in the dataset is as follows: 14\% in 1a, 18\% in 1b, 10\% in 2a, 24\% in 2b, 12\% in 3a, 7\% in 3b, and 15\% in 4. In Section \ref{subsec:dataIR}, we compare this dataset distribution with the rest of the medical literature.}

The Oncology Guidelines mention publications as citations, which include the authors names, publication year, and publication title. This information is not sufficient for automatic LoE classification, which additionally requires some of the methodology, interventions, and clinical outcomes. This information can only be found in publication abstracts or full texts. Therefore, we leverage the PubMed API\footnote{\url{https://pubmed.ncbi.nlm.nih.gov/}} to enrich the initial dataset with abstracts and PubMed IDs. 
 
The average word count in the abstracts is 263 (SD=97), slightly above the typical range for medical articles~\citep{andrade2011write}. The prevalence of longer abstracts can be attributed to the frequent use of structured abstract formats within the medical literature~\citep{hartley2004current}. Notably, we observe a positive correlation between the abstract length and the LoE classification: publications with higher evidence levels tend to have longer abstracts (e.g.\ LoE 1a with a mean of 325 words (SD=163) than those with lower levels (LoE 3b and 4 with a mean of 233 words (SD=71)).

We split this data into a training dataset containing 1690 instances (60\%) and a validation and testing dataset containing 563 instances (20\%) each, ensuring a stratified representation across all classes. 

\subsection{Experimental Setup}

For the task of LoE classification, we focus on fine-tuning PubMedBERT~\citep{gu2021domain}. PubMedBERT is a natural choice for this domain-specific classification task as it is a transformer-based model pre-trained using abstracts sourced directly from PubMed. Its efficacy has been well-established:  It currently holds the top score on the Biomedical Language Understanding and Reasoning Benchmark~\citep{gu2021domain}, it excels in accurately interpreting the unique terminologies and context of biomedical texts, and it is proficient in handling the complexities of biomedical literature. The model is fine-tuned using the training set and hyperparameters are optimized using the validation set. 
%
%
%
%
We develop the following classifiers: 

\paragraph{\textbf{Random Forest (RF)}}
RF serves as our baseline. It is trained on the training set for multi-class classification. We use TF-IDF vectorization and chi-squared feature selection, and K-Fold cross-validation using the validation dataset, evaluating its performance with the macro-F1 score. 

\paragraph{\textbf{Multi-Class-PubMedBERT}}
This classifier is directly fine-tuned on the training set to classify texts into specific LoE classes, with the macro-F1 score as the evaluation matrix.

\paragraph{\textbf{Reg-PubMedBERT}}
This is a regression approach, which assigns numeric values to LoE classes. PubMedBERT is fine-tuned to predict these values, by mapping different LoEs (1a, 1b, 2a, 2b, 3a, 3b, 4) to the numeric values (0, 1, 2, 3, 4, 5, 6). We used root-mean-square error (RMSE) for evaluation. To align the model's predictions with the original LoE classes and to facilitate comparison with other classifiers using the F1 matrix, we mapped the predicted value to the nearest integer value and then used the same map to get predictions back to their corresponding LoE classes.

\paragraph{\textbf{Multi-Label-PubMedBERT}}
This classifier incorporates the multi-label classification approach, i.e.\  we transform the LoE categorization into a set of binary labels. Each label corresponds to a specific LoE class, effectively converting the problem into a multi-label classification task. This version enabled PubMedBERT to predict multiple labels simultaneously, accommodating the scenario where only one of the labels should be true while others are false. By modelling the LoE classification as a multi-label task, we aim to capture potential overlap between LoE classes and assess the model's capacity to handle such nuances by looking at the prediction list that might contain multiple levels of evidence. For proper evaluation, we assigned the highest confidence value when multiple positive predictions. 

\paragraph{\textbf{Ensemble Majority Vote}}

Ensemble methods are a well-established technique in classification that capitalizes on the strengths of diverse classifiers to enhance prediction accuracy and generalization~\citep{polikar2012ensemble}. We employed an Ensemble Majority Vote strategy, combining the strengths of the three PubMedBERT models (Multi-Class, Reg, and Multi-Label). This approach used majority voting to aggregate predictions from each model, enhancing the overall classification accuracy and robustness~\citep{zhou2021ensemble,dang-etal-2020-ensemble}.

\subsection{Classifier Evaluation}
We evaluate our LoE document classifiers using Macro F1 score, RMSE, and Confusion matrices. 
\added{RMSE treats the LoE classification as a regression task by mapping each LoE category to a corresponding integer. The RMSE score reflects the average squared difference between the predicted and true LoE values, providing important insight into how closely the model captures the ordinal nature of the LoE hierarchy. Lower RMSE values indicate that the model's predictions are closer to the true LoE, particularly emphasizing the reduced impact of misclassifications to adjacent levels.}

\subsubsection{Individual Classifiers Performance}
Table~\ref{tab:modelsScores} summarizes the performance of each classifier on the test dataset.

\begin{table}[htp!]
    \centering
    \begin{tabular}{|c|c|c|} \hline 
         Model                  & F1 score  & RMSE  \\ \hline 
         Random Forest (RF)         & 0.59      &  1.30 \\ \hline 
         Multi-Class-PubMedBERT & 0.78      &  0.90 \\ \hline 
         Reg-PubMedBERT         & 0.74      &  0.69 \\ \hline 
         Multi-Label-PubMedBERT & 0.79      & $\;$0.90\textsuperscript{*} \\ \hline 
         Majority voting        & \textbf{0.83}  & \textbf{0.65} \\ \hline
    \end{tabular}
    \\
    \centering
    \caption{Level of Evidence Classifiers Performance on our test set. Macro F1 Score.} 
      \noindent{\footnotesize{\textsuperscript{*} By considering the label of the highest confidence score as predicted class}}
  \label{tab:modelsScores}
\end{table}

\begin{figure}[hbt!]
\centering
\subfloat[Random Forest as a baseline]{\includegraphics[width=0.4\textwidth]{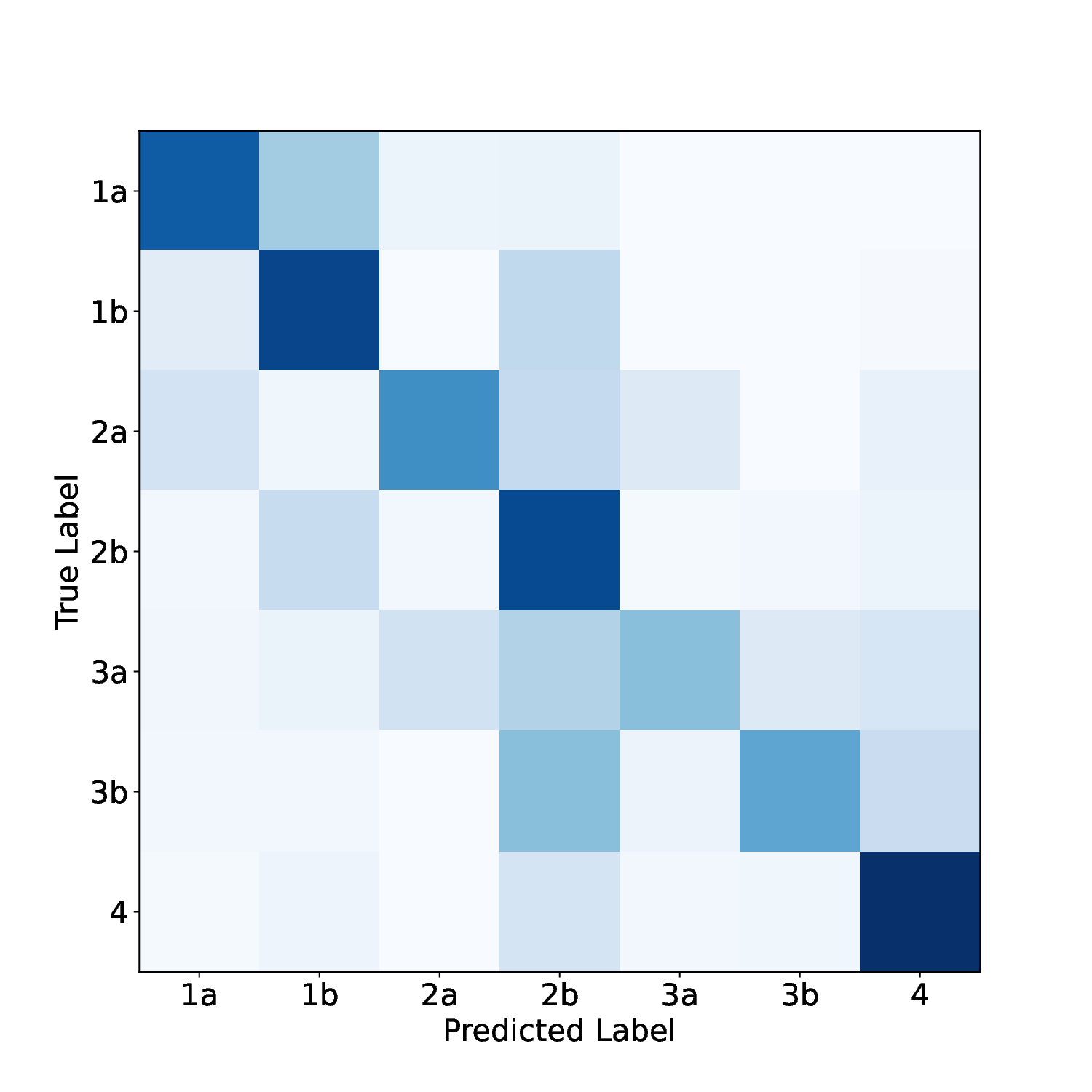}
\label{fig:randomforest}}
\hfill
\subfloat[Multi-Class-PubMedBERT]{\includegraphics[width=0.4\textwidth]{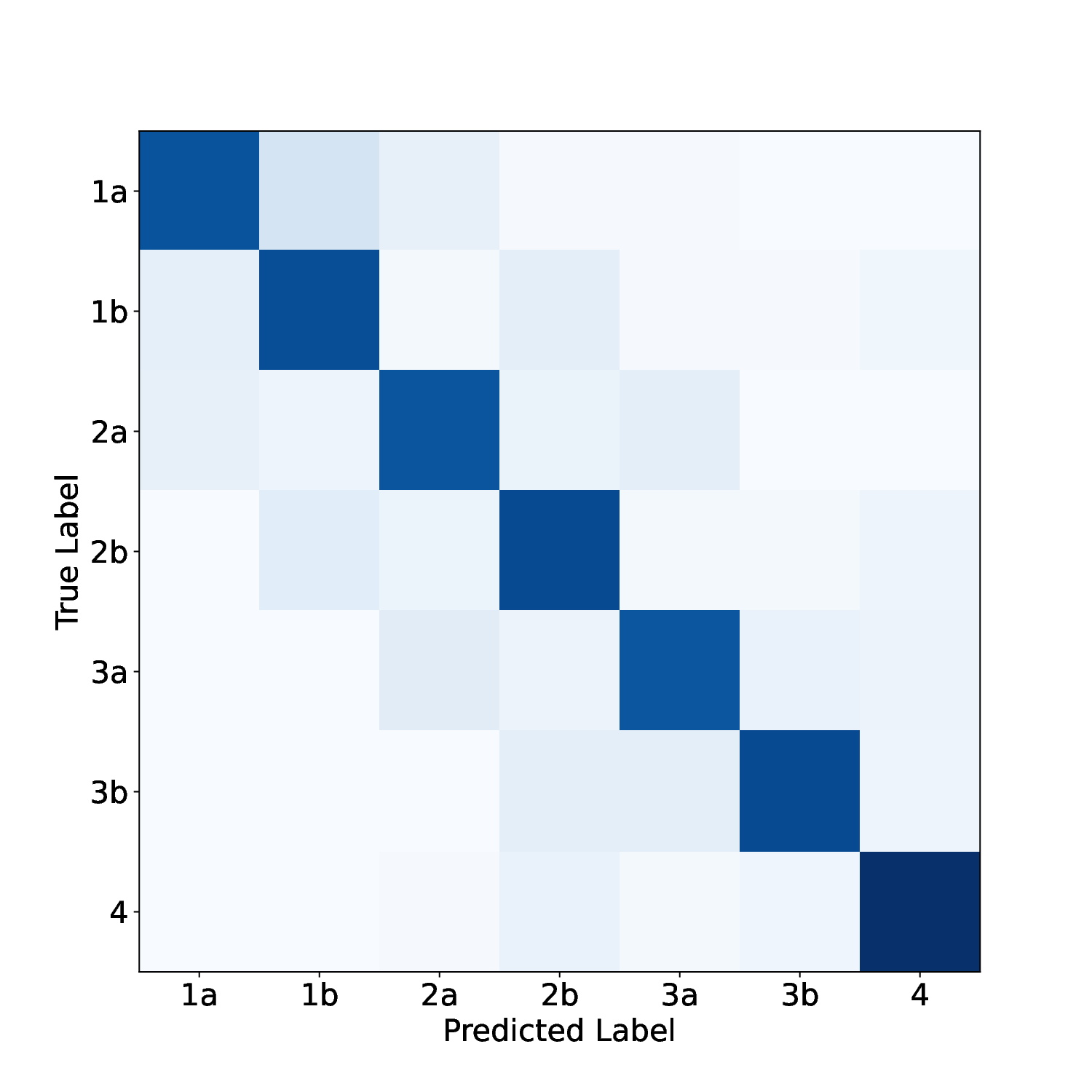}
\label{fig:multiclass}}
\hfill
\subfloat[Reg-PubMedBERT]{\includegraphics[width=0.4\textwidth]{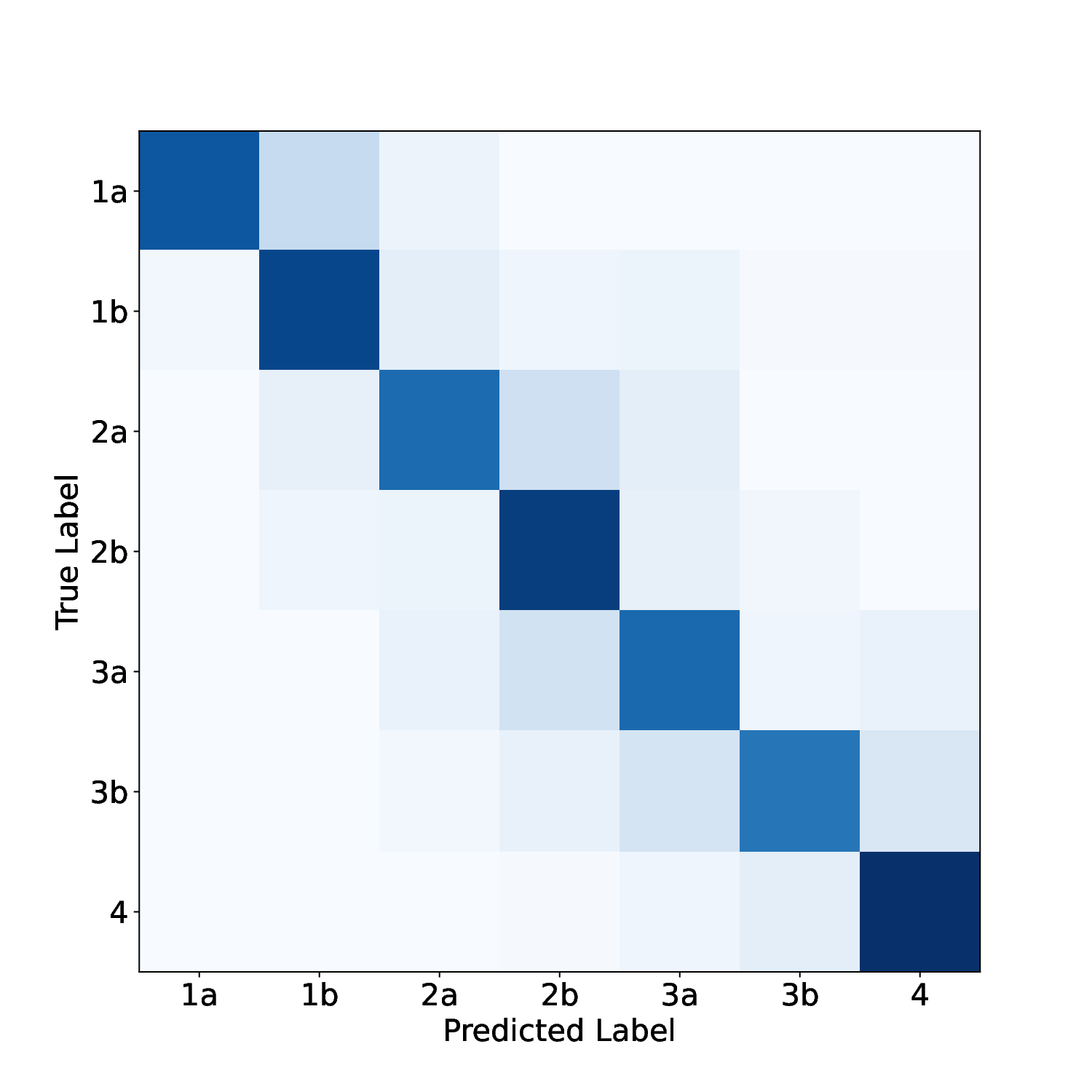}
\label{fig:regression}}
\hfill
\subfloat[Majority voting]{\includegraphics[width=0.4\textwidth]{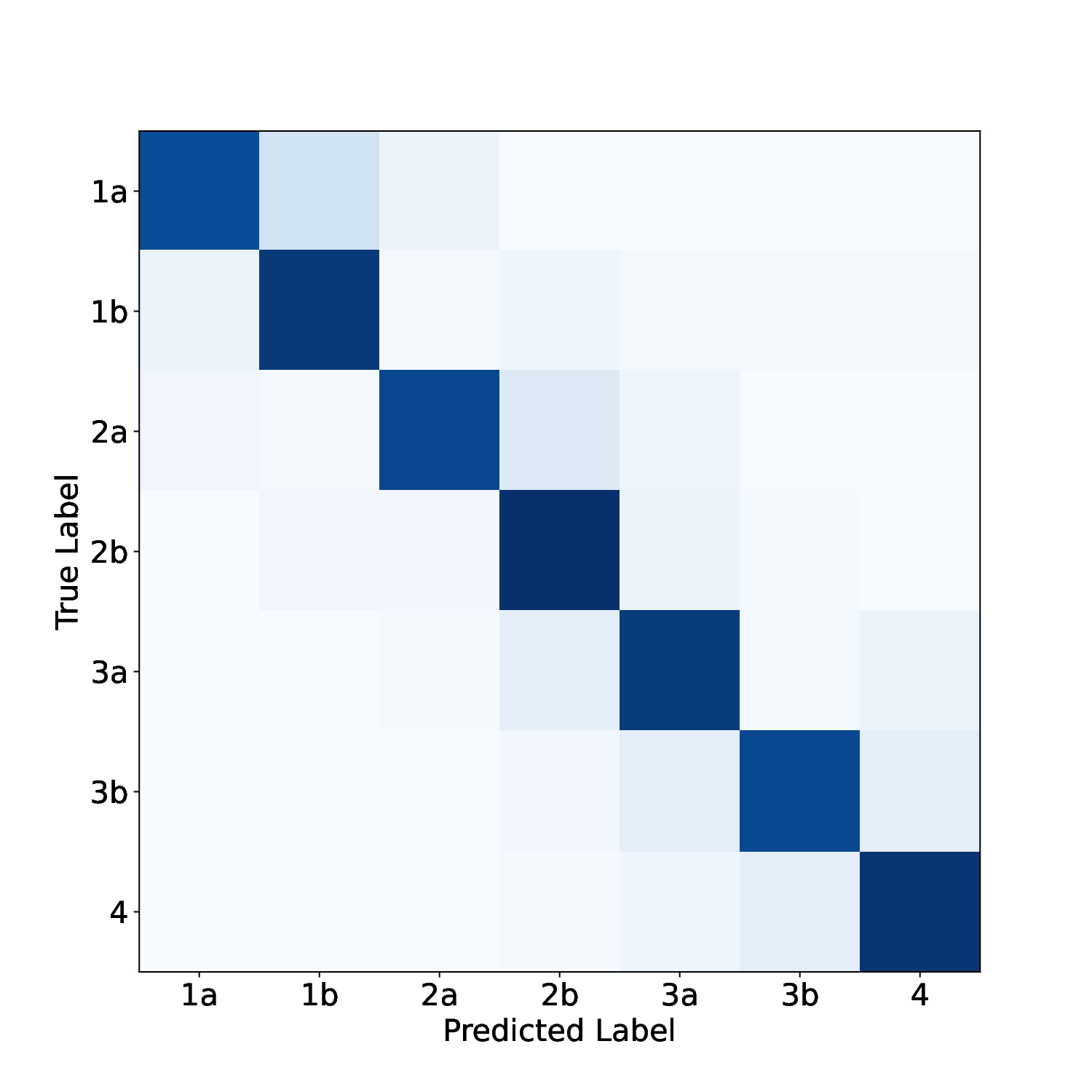}
\label{fig:majorityvoting}}
\caption{Confusion Matrices using the test set per model.}
\label{fig:confusionMatrix}
\end{figure}

\paragraph{\textbf{RF Baseline}}
The RF model's performance with a macro-F1 score of 0.59 and an RMSE of 1.30 did not surpass the deep learning models' results. Nevertheless, the RF model shows robustness in effectively handling the challenges of multi-class LoE classification. Analyzing the confusion matrix in Figure~\ref{fig:randomforest}, we see that the misclassifications are rather scattered, they are not clustered in any particular class or in neighboring classes. 

\paragraph{\textbf{Multi-Class-PubMedBERT}}
Multi-Class-PubMedBERT scored 0.78 in F1 (+0.19 compared with baseline) and 0.90 in RMSE, showing effectiveness in multi-class categorization. However, after we analysed misclassification in Figure~\ref{fig:multiclass}, we found that the model has some difficulties distinguishing closely related LoE classes. This suggests considering the problem as a regression task, since misclassification with neighbor classes is not as bad as assigning distant classes such as replacing 1a with 4.

\paragraph{\textbf{Reg-PubMedBERT}}
exhibited strengths in capturing the ordered nature of LoE with an F1 score of 0.74 and the second-best RMSE of 0.69, indicating proficiency in differentiating between levels. This makes misclassified documents closer to the true labels, which is reflected in the smaller RMSE and highlighted in Figure~\ref{fig:regression}.  This makes misclassification to neighboring classes less harmful than assigning far classes. 

\paragraph{\textbf{Multi-Label-PubMedBERT}}
The model performed best among individual classifiers with an F1 score of 0.79, adeptly handling documents with multiple LoE categories. A closer qualitative examination of this model's performance revealed that some documents were assigned into multiple LoE classes. This is a well known phenomenon, which was explored in the work of~\citet{murad2016new} bringing into question the clear demarcation between the evidence levels of the EBM pyramid. Instead, a nuanced perspective on LoEs has been proposed to align with the flexibility of multi-label classification as demonstrated by Multi-Label-PubMedBERT.

\subsubsection{Ensemble Majority Vote Performance}
The Ensemble Majority Vote method combines the predictions of all three PubMedBERT models and demonstrates the best performance. It scores highest in F1 (0.83) and achieves an RMSE of 0.65, indicating its effectiveness in accurately categorizing medical literature by LoE. This result emphasizes the significant role of collaborative intelligence in enhancing classification outcomes. It also benefited from the power of the regression model, where misclassification resulted in neighboring classes as shown in Figure~\ref{fig:majorityvoting} and the smallest RMSE.  

\subsubsection{Statistical Significance Analysis}
We performed a statistical significance analysis on our machine learning models using a paired t-test. After applying Bonferroni correction ($\alpha$ = 0.05/10), we found that all 
deep learning models significantly outperformed the Random Forest baseline, indicating their effectiveness in LoE classification. However, no significant performance differences were observed among the deep learning models themselves, highlighting their comparable efficacy in evidence-based classification.

\subsubsection{Identifying Significant Terms}
We utilized the LIME (Local Interpretable Model-Agnostic Explanations) explainer~\citep{lime2016why} to identify key terms influencing our model's predictions for different Levels of Evidence (LoE) categories. This method provides insights by aggregating term scores, helping us to determine significant terms for each LoE level. Such an approach enhanced the interpretability and transparency of our model, highlighting LoE-specific terms in the analyzed documents.


Table~\ref{tab:wordcloud} presents the top 10 contributing terms across the LoE levels in the test set. The results highlighted that our model was able to identify discriminating terms for each class. Moreover, we discovered common terms shared across multiple levels, such as ``systematic review" in 1a (systematic reviews of RCTs), 2a (systematic reviews of cohort studies), and 3a (systematic reviews of case-control studies), and ``RCT" in 1a and 1b (individual RCTs). Additionally, some less expected terms, like ``risk" in 2a, 2b (individual cohort studies), 3a, and 3b (individual case-control studies), and ``accuracy study" in 1a, 2a, and 3a (pertaining to Diagnostic Test Accuracy studies), emerged as significant classifiers. Interestingly, a specific therapy (``acupuncture") only occurs among the terms of level 4, possibly indicating the lack of stronger evidence for this method.

\begin{table}[!t]
\centering
\hspace*{-0.4cm}\resizebox{0.95\textwidth}{!}{
    \begin{minipage}{\linewidth}
        \begin{tabular}{|l|l|l|l|l|l|l|l|}
\hline
\multicolumn{2}{|c|}{1a}                       & \multicolumn{2}{c|}{1b}                      & \multicolumn{2}{c|}{2a}                      & \multicolumn{2}{c|}{2b} \\ \hline
term             & $\!$score$\!\!$ & term            & $\!$score$\!\!$ & term            & $\!$score$\!\!$ & term          & $\!$score$\!\!$        \\ \hline
accur predict    & 2.11  &  achiev complet$\!\!$   & 1.92  &  cohort studi     & 1.30  &  cohort studi    & 1.62   \\ \hline
accur stage       & 1.85  &  achiev patient$\!\!$   & 1.91  &  accuraci detect  & 1.14  &  accrual         & 1.42  \\ \hline
accuraci respect$\!\!$  & 1.72  &  activ control    & 1.58  &  systemat review$\!\!$  & 1.09  &  acquisit        & 1.14  \\ \hline
rct               & 1.42  &  activ intervent$\!\!$  & 1.56  &  meta analysi     & 1.02  &  accept          & 1.11  \\ \hline
meta analysi      & 1.31  &  activ surveil    & 1.25  &  exposur          & 0.98  &  access          & 1.08  \\ \hline
systemat review$\!\!$   & 1.30  &  rct              & 1.21  &  longitudin       & 0.95  &  accru           & 1.01  \\ \hline
accuraci studi    & 1.17  &  control set      & 1.12  &  access           & 0.74  &  longitudin      & 0.89  \\ \hline
accuraci clinic   & 1.16  &  acut delay       & 0.98  &  accur stage      & 0.73  &  risk            & 0.61  \\ \hline
achiev            & 1.15  &  acut             & 0.79  &  accuraci studi   & 0.64  &  administr       & 0.21  \\ \hline
activ treatment   & 1.02  &  adjuv            & 0.71  &  risk             & 0.59  &  affect patient$\!\!$  & 0.14  \\ \hline
\noalign{\smallskip}
        \end{tabular}
    \end{minipage}
}
    \\
\hspace*{+0.9cm}\resizebox{0.95\textwidth}{!}{
    \begin{minipage}{\linewidth}
        \begin{tabular}{|l|l|l|l|l|l|}
\noalign{\smallskip}
\hline
\multicolumn{2}{|c|}{3a}                       & \multicolumn{2}{c|}{3b}                      & \multicolumn{2}{c|}{4}                            \\ \hline
term            & score & term           & score &term                 & score \\ \hline
systemat review   & 1.24  &  case control     & 1.60  &  small sampl           & 1.69  \\ \hline
epidemiolog       & 1.21  &  case definit     & 1.41  &  preliminari evid      & 1.32  \\ \hline
case definit      & 1.17  &  exposur          & 1.02  &  exploratori research  & 0.99  \\ \hline
abnorm            & 1.12  &  risk             & 0.49  &  uncontrol studi       & 0.98  \\ \hline
exposur           & 1.11  &  advers reaction  & 0.31  &  acupunctur treatment  & 0.68  \\ \hline
absent            & 0.98  &  affect patient   & 0.30  &  patient characterist  & 0.60  \\ \hline
accuraci respect  & 0.88  &  age              & 0.29  &  acupunctur effect     & 0.51  \\ \hline
accuraci studi    & 0.71  &  age diagnosi     & 0.23  &  analysi reveal        & 0.22  \\ \hline
risk              & 0.64  &  advers effect    & 0.19  &  analysi identifi      & 0.22  \\ \hline
accur stage       & 0.51  &  affect surviv    & 0.10  &  affect                & 0.13  \\ \hline
        \end{tabular}
    \end{minipage} 
}
    \caption{Significant Terms in the Level of Evidence Classifier.}
\label{tab:wordcloud}
\end{table}

\section{Levels of Evidence as a filter in medical IR} \label{sec:loeFilters}

\added{For the retrieval experiments, the 7-class LoE model was simplified into a 4-class setup by grouping related evidence levels. This decision reflects real-world usage patterns where users often prioritize broader evidence categories, such as high-quality studies (e.g., systematic reviews and RCTs) or intermediate-level evidence (e.g., cohort and case-control studies). This reduction not only simplifies classification, but also improves retrieval effectiveness without compromising performance.}

In this experiment, we investigate the benefit of LoE classification for the IR of medical publications using TREC Precision Medicine (PM) datasets from 2017 to 2019~\citep{roberts2017overview,roberts2018overview,roberts2019overview}. 

\subsection{Data} \label{subsec:dataIR}
The TREC PM datasets, sourced from the Medline collection\footnote{\url{https://www.nlm.nih.gov/medline/medline_overview.html}}, consist of over 26 million research article abstracts accessible via PubMed and designed to enhance biomedical IR. Topics/queries were constructed based on disease and gene fields from the dataset, omitting demographic data to focus specifically on abstract retrieval.
Relevance judgements were performed by expert assessors on a scale of `not relevant (0)', `partially relevant (1)', and `definitely relevant (2)', based on alignment with a given topics~\citep{roberts2017overview}. The criteria for relevance did not include the LoE of the documents.

\begin{figure}
\centering
\includegraphics[width=0.8\textwidth]{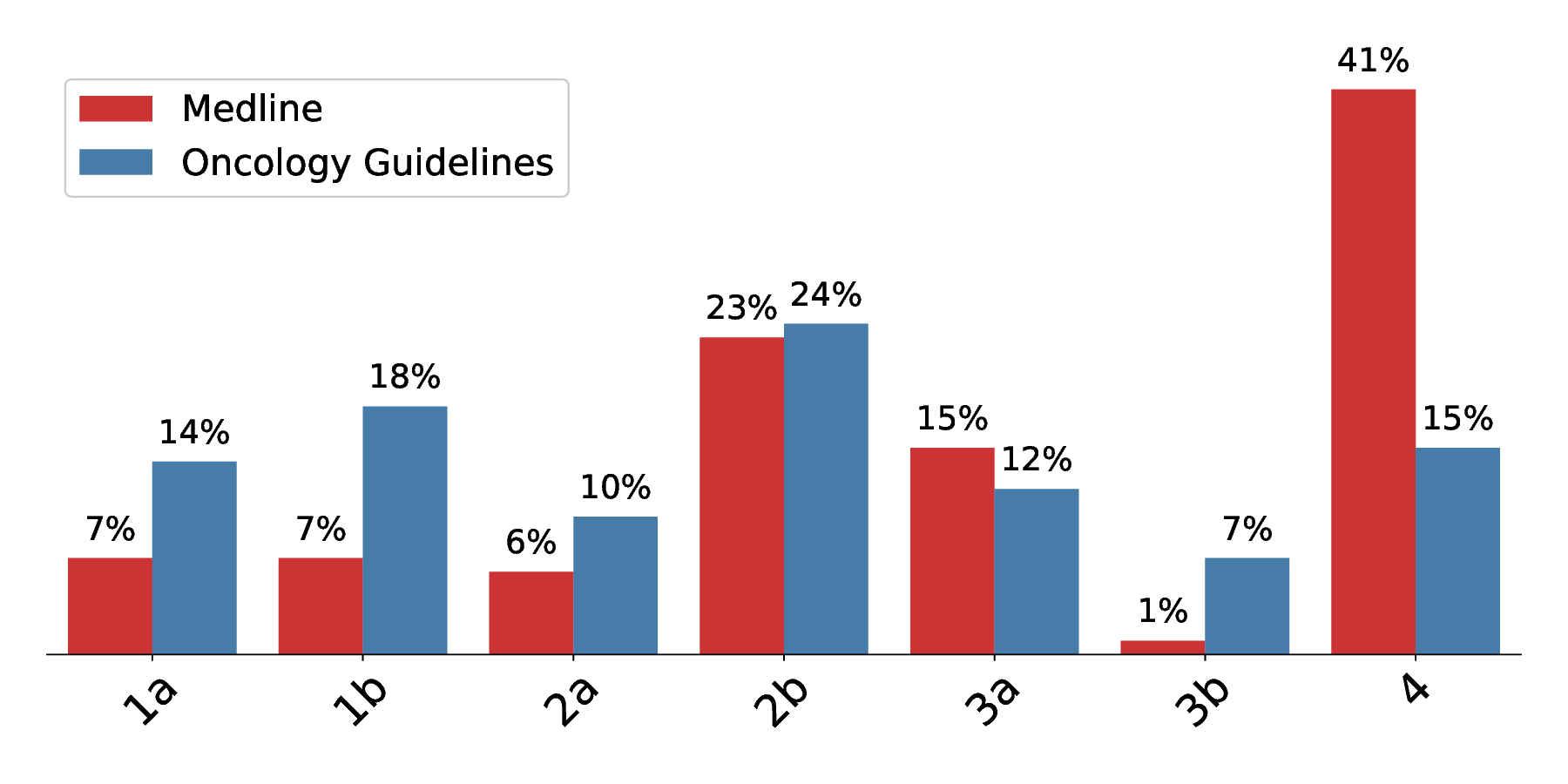}
\caption{The distribution of LoE Classes in the Medline Dataset and Oncology Guidelines (Classifier Dataset).}
\label{fig:distribution}
\end{figure}

We categorize each abstract in the Medline collection into its respective LoE category using our ensemble classifier. Figure~\ref{fig:distribution} shows the distribution of LoE classes in Medline data. Most frequent are Level 4 documents (41\% of the collection), which require the smallest empirical basis. The highest LoE 1a and 1b each represent only 7\% of the documents. \replaced{This unbalanced distribution reflects the inherent nature of the biomedical literature, where expert opinion and hypothesis-generating studies far outnumber high-evidence clinical research, which is usually conducted after several studies have confirmed the same observations.}{This distribution highlights the predominance of lower-evidence articles in medical literature and underscores the importance of our approach in focusing on evidence quality in IR.}

\added{In contrast, the oncology guidelines exhibit a reversed pattern, with  only 15\% of low evidence documents, suggesting a higher balance ratio. This difference is due to the selection process during the formulation of new clinical guidelines, where publications with higher evidence are prioritized. }

\subsection{Experimental Setup}
In this subsection, we present the IR methods used for performing this task and the evaluation metric.
As a core retrieval algorithm, we use BM25 \citep{robertson2009probabilistic}, which is widely used in IR for scoring and ranking documents based on their relevance to a user query. It is a probabilistic-based approach that builds on the classic TF-IDF (Term Frequency-Inverse Document Frequency) model, refining it by incorporating factors like term frequency saturation and document length normalization. The algorithm calculates a score for each document by considering how frequently the query terms appear in the document, adjusting for the overall document length and the rarity of the terms across the entire corpus. BM25 is particularly valued for its ability to effectively balance term frequency and inverse document frequency, making it one of the most robust and popular methods for ranking search results.

\added{In the retrieval experiment, we first indexed the entire Medline collection (Sec. \ref{subsec:dataIR}) alongside their assigned LoE classes. The BM25 algorithm, parameterized with $K_1=1.2$ and $b=0.75$ as recommended in \citep{connelly2019practical}, was then applied to retrieve and rank documents based exclusively on textual relevance (abstracts and titles) to the query, without integrating LoE into the ranking process. This methodology ensures a fair comparison with the baseline.}

\subsubsection{Retrieval Methods}\label{subsec:IRMethods}
Our experiment utilizes the BM25 retrieval method applied to documents of all LoE classes (`All') as a baseline for our IR process. The impact of LoE classification is tested by filtering the documents based on their LoE as follows:
\begin{itemize}
    \item \textit{LoE3+:} LoE categories 3b to 1a, i.e.\ case-control studies or higher LoE.
    \item \textit{LoE2+:} LoE categories 2b to 1a, i.e.\ cohort studies or higher LoE.
    \item \textit{LoE1:} LoE categories 1a and 1b, i.e.\ RCTs only.
\end{itemize}




\subsubsection{Evaluation Metric}
The performance of each model's effectiveness was assessed using infNDCG, R-Prec, and P@10 matrices, as these are the official matrices used to report on the datasets. Also, we report the ``Normalized discounted cumulative gain @10'' (NDCG@10) metric as our core metric~\citep{DBLP:conf/sigir/JarvelinK00}. This measure allows for considering relevance grades $0\ldots n$, where, in our case, irrelevant documents receive a score of 0, while partially relevant and definitely relevant documents receive scores of 1 and 2, respectively. For a ranked document list, let $r_j$ denote the relevance grade of the document at rank $j$. Then the (unnormalized) discounted cumulative gain for a ranked list of length $k$ is defined as: 
   \[
   DCG(k) = \sum_{j=1}^k\frac{r_j}{max(1,\log_b(j))}.
   \]

\noindent
With the denominator in the summation elements, DCG simulates a stochastic user stopping behavior, where not every user checks all documents up to the final rank $k$,  but  some users might stop at earlier ranks; the fraction of users reaching a certain rank is  controlled by the logarithm base $b$ (usually chosen as $b=2$).
As the DCG values for a query depend heavily on the number of relevant documents in the collection, they are normalized by comparing them with the value $DCG_{opt} (k)$ of the optimum retrieval result (i.e.\ ranking documents by decreasing relevance grades), thus arriving at the normalized discounted cumulative gain:
\[
N\!DCG@k = {DCG(k)}/{DCG_{opt}(k)}.
\]

\noindent
Besides incorporating a fairly realistic user stopping behavior and being one of the few retrieval metrics considering different relevance grades, NDCG also has a nice theoretic property: \citep{DBLP:journals/access/FerranteFF21} showed that NDCG comes closest to an interval scale (which is a requirement for computing means and effect sizes), while other popular measures with stochastic stopping behavior (like  average precision or rank-biased precision) clearly violate this property.


\subsection{Results}

As shown Table~\ref{tab1} using LoE to filter out document set to be searched improves the retrieval effectiveness as measured by NDCG@10 score. The retrieval of RCT documents with highest LoEs is the most successful. Moreover, there is a clear trend in improving NDCG when the minimum LoE is increased. For all three collections, the strictest filter \added{(LoE1 with only 14\% of the collection)} outperformed all other methods, with substantial NDCG improvements (0.08 $\ldots$ 0.11) over the baseline.  As we are re-using a test collection, performing statistical tests here would contradict statistical testing theory~\citep{Fuhr:17b}. Instead, we give the effect sizes, which indicate substantial improvements over the baseline.

Moreover, as shown in Table~\ref{tab:officalresults}, our LoE1 model improved the performance of the baseline on all matrices. It also outperformed each of the best-reported runs on infNDCG matrix and provided comparable results on R-Prec\footnote{Note that these are pessimistic estimates, as unjudged documents only retrieved by our method are treated as irrelevant}. In addition, the retrieval quality of our method is accompanied with the guarantee of returning only  documents of the highest evidence. \added{On the other hand, as the results for P@10 show, LoE seems to be too strict when the user is  looking at all 10 top-ranking documents.}

\begin{table}[th!]
    \centering
    \begin{tabular}{|l|r||l|l|l|} 
        \noalign{\smallskip}
        \hline 
        Exp./Year & \multicolumn{1}{c||}{size*} & 2017 & 2018 & 2019 \\ \hline 
        \textit{All} & 100\% &0.46 & 0.59 & 0.54 \\ \hline
        \textit{LoE3+}& 59\% &0.48 (0.02)** & 0.60 (0.01) &  0.57 (0.03) \\ \hline 
        \textit{LoE2+} & 43\% & 0.49 (0.03) & 0.64 (0.05) &  0.58 (0.04) \\ \hline
        \textit{LoE1}& 14\% & \textbf{0.54} (0.08) & \textbf{0.69} (0.10) &
        \textbf{0.65} (0.11) \\ \hline
    \end{tabular}
    \caption{Models' NDCG@10 performance on TREC PM datasets}
    \vspace{1mm}
    \noindent{\footnotesize{\textsuperscript{*} Size denotes the percentage of the collection that was considered in retrieval. \\
    \footnotesize{\textsuperscript{**} Numbers in parentheses show the effect size when comparing with the baseline ``All''.}}}
    \label{tab1}
    \vspace{-2mm}
\end{table}

\begin{table}[th!]
    \centering
    \begin{tabular}{|l|l|l|l|} \hline 
        Exp./Year      & 2017              & 2018              & 2019           \\ \hline 
        \textit{All}   & 0.43~/~0.27~/~0.52    & 0.50~/~0.32~/~0.58    & 0.47~/~0.30~/~0.57 \\ \hline
        \textit{LoE3+} & 0.45~/~0.28~/~0.54    & 0.52~/~0.34~/~0.60    & 0.50~/~0.31~/~0.58 \\ \hline 
        \textit{LoE2+} & 0.47~/~0.28~/~0.54    & 0.55~/~0.36~/~0.61    & 0.52~/~0.31~/~0.61 \\ \hline
        \textit{LoE1}  & \textbf{0.52}~/~\textbf{0.30}~/~0.55 & \textbf{0.57}~/~\textbf{0.38}~/~0.61 & \textbf{0.58}~/~0.34~/~0.61 \\ \hline
       \textit{Top run}& 0.46~/~\textbf{0.30}~/~\textbf{0.64} & 0.56~/~0.37~/~\textbf{0.71} & \textbf{0.58}~/~\textbf{0.36}~/~\textbf{0.65} \\ \hline
    \end{tabular}
    \caption{Models' InfNDCG/R-Prec/P@10 performance on TREC PM datasets.*}
    \vspace{1mm}
    \noindent{\footnotesize{\textsuperscript{*} {Best reported runs per matrix, meaning the model performing best on P@10 is not the same as the model performing best on infNDCG.}}}
\label{tab:officalresults}
    \vspace{-2mm}
\end{table}

\section{Discussion} \label{sec:discussion}

In this paper, we have effectively demonstrated the automated application of the LoE framework for improving the retrieval of relevant medical publications. Our approach, leveraging fine-tuned PubMedBERT models, has proven adept at classifying medical publications based on their LoE with a high degree of accuracy (macro F1 = 0.83). This advancement addresses a significant gap in existing literature, where previous studies have largely focused on specific evidence levels, particularly RCTs and their systematic reviews. The higher transparency of our approach gives users full control over the LoE of the documents returned. Moreover, the method investigated here could be directly integrated into the existing PubMed search engine, by simply adding estimated LoE as an additional document attribute that can be referred to in the query.


A key finding of our work is the effect of LoE filtering in directing attention towards the most reliable 14\% of documents, while enhancing retrieval quality at the same time. This aspect is particularly crucial in the medical domain, where accessing accurate and high-quality information rapidly can make a pivotal difference in patient care and medical research. On the other hand, LoE2 or LoE3 papers may also be searched for in case there are no relevant answers in the top level, e.g.\ when the user is interested in more recent methods for which higher level studies are not available yet. \added{Therefore, we acknowledge that clinical decision-making often requires synthesizing multiple sources across different LoE levels.}



In our study, the LoE1 model outperformed the best-reported runs on the three datasets \citep{roberts2017overview,roberts2018overview,roberts2019overview} in terms of infNDCG and provided comparable results in R-Prec matrix. This demonstrated the effectiveness of using LoE as a filter in medical IR, improving the relevance and reliability of retrieved documents. These improvements over integrating the LoE filter in the BM25 baseline suggest that these benefits could extend to the other stronger baselines.

Although our study shows the potential of using LoE in Medline, one limitation that needs to be considered is the potential bias from using the oncology guideline dataset for training the classifiers. Medline collection contains publications where LoE can not be applied, such as bioinformatics. To apply it in real-world applications, we could introduce a new class, "others", where the model confidence score is below the seine threshold or when multiple positive labels are in the multi-label classifier. 

\added{Moverover, the LoE framework prioritizes study design rigor but does not assess study quality (e.g., risk of bias). Future work should integrate tools like GRADE~\citep{guyatt2009grade} or Cochrane’s risk of bias assessment to enhance reliability. This requires expanding the research article and analyzing the full text rather than the title and abstract, which are enough for assigning LoE.}

\replaced{In our recently published user study~\citep{JCDL:UserStudy2024}, we present findings from an evaluation with medical professionals testing a clinical search engine that integrates LoE classification with biomedical concepts as a semantic layer~\citep{frihat2025bioconcepts}.

The results demonstrated strong user engagement with LoE: 93\% of participants reported prior familiarity with LoE frameworks, and 85\% actively filtered search results based on high LoE levels, noting that this feature facilitated their ability to prioritize high-quality evidence. Their feedback also highlighted the added value of biomedical concept extraction (e.g., gene-disease relationships) in contextualizing evidence.}{In an ongoing user study with medical professionals we investigate if these experimental enhancements also lead into measurable benefits in real-world clinical settings. By engaging real users in this research, we intend to validate the practical utility of our LoE integrated system in enhancing search experiences and outcomes.}


\section{Conclusion} \label{sec:conclusion}

Our research addresses the challenge faced by current search engines in identifying significant, evidence-backed medical publications. Although relevant and widely used in evidence-based medical practice, the LoE framework has not yet been fully automatised and tested for medical IR. We introduce a classification model for tagging medical research abstracts with LoE levels and demonstrate that a vast number of medical publications without LoE tags can be successfully and fully automatically enriched with this crucial information. Our retrieval results confirm that LoE is an effective filter that improves results in a fully automatic retrieval scenario. These results suggest that our LoE based approach to medical IR is a viable and robust tool to evidence-based medical practice, which can facilitate and improve medical decision-making, leading to better patient care. \added{However, effective decision-making often requires synthesizing multiple studies and integrating clinical practice guidelines, which remains an important area for future work.}




\acks{%
This work was funded by a PhD grant from the DFG Research Training Group 2535 Knowledge- and data-based personalization of medicine at the point of care (WisPerMed), University of Duisburg{-}Essen, Germany.  We also acknowledge support by the Open Access Publication Fund of the University of Duisburg{-}Essen.
}

\vskip 0.2in

\begin{thebibliography}{40}
\providecommand{\natexlab}[1]{#1}
\providecommand{\url}[1]{\texttt{#1}}
\expandafter\ifx\csname urlstyle\endcsname\relax
  \providecommand{\doi}[1]{doi: #1}\else
  \providecommand{\doi}{doi: \begingroup \urlstyle{rm}\Url}\fi

\bibitem[Andrade(2011)]{andrade2011write}
Chittaranjan Andrade.
\newblock How to write a good abstract for a scientific paper or conference
  presentation.
\newblock \emph{Indian Journal of Psychiatry}, 53\penalty0 (2):\penalty0 172,
  2011.

\bibitem[Arno et~al.(2022)Arno, Thomas, Wallace, Marshall, McKenzie, and
  Elliott]{arno2022accuracy}
Anneliese Arno, James Thomas, Byron Wallace, Iain~J Marshall, Joanne~E
  McKenzie, and Julian~H Elliott.
\newblock Accuracy and efficiency of machine learning--assisted risk-of-bias
  assessments in ``real-world'' systematic reviews: A noninferiority
  randomized controlled trial.
\newblock \emph{Annals of Internal Medicine}, 175\penalty0 (7):\penalty0
  1001--1009, 2022.

\bibitem[Borawski et~al.(2007)Borawski, Norris, Fesperman, Vieweg, Preminger,
  and Dahm]{borawski2007levels}
Kristy~M Borawski, Regina~D Norris, Susan~F Fesperman, Johannes Vieweg, Glenn~M
  Preminger, and Philipp Dahm.
\newblock Levels of evidence in the urological literature.
\newblock \emph{The Journal of Urology}, 178\penalty0 (4):\penalty0 1429--1433,
  2007.

\bibitem[Burns et~al.(2011)Burns, Rohrich, and Chung]{burns2011levels}
Patricia~B Burns, Rod~J Rohrich, and Kevin~C.Chung.
\newblock The levels of evidence and their role in evidence-based medicine.
\newblock \emph{Plastic and Reconstructive Surgery}, 128\penalty0 (1):\penalty0
  305, 2011.

\bibitem[Cohen et~al.(2021)Cohen, Schneider, Fu, McDonagh, Das, Holt, and
  Smalheiser]{cohen2021fifty}
Aaron~M Cohen, Jodi Schneider, Yuanxi Fu, Marian~S McDonagh, Prerna Das,
  Arthur~W Holt, and Neil~R Smalheiser.
\newblock Fifty ways to tag your pubtypes: Multi-tagger, a set of probabilistic
  publication type and study design taggers to support biomedical indexing and
  evidence-based medicine.
\newblock \emph{medRxiv}, pages 2021--07, 2021.

\bibitem[Connelly(2019)]{connelly2019practical}
Shane Connelly.
\newblock Practical BM25 -- Part 3: Considerations for picking b and k1 in
  Elasticsearch, 2019.
\newblock URL
  \url{https://www.elastic.co/blog/practical-bm25-part-3-considerations-for-picking-b-and-k1-in-elasticsearch}.

\bibitem[Dang et~al.(2020)Dang, Lee, Henry, and
  Uzuner]{dang-etal-2020-ensemble}
Huong Dang, Kahyun Lee, Sam Henry, and {\"O}zlem Uzuner.
\newblock Ensemble {BERT} for classifying medication-mentioning tweets.
\newblock In \emph{Proceedings of the Fifth Social Media Mining for Health
  Applications Workshop {\&} Shared Task}, pages 37--41, Barcelona, Spain,
  2020. Association for Computational Linguistics.
\newblock URL \url{https://aclanthology.org/2020.smm4h-1.5}.

\bibitem[Desai et~al.(2019)Desai, Camp, and Krych]{desai2019hierarchy}
Vishal~S Desai, Christopher~L Camp, and Aaron~J Krych.
\newblock What is the hierarchy of clinical evidence?
\newblock \emph{Basic Methods Handbook for Clinical Orthopaedic Research: A
  Practical Guide and Case Based Research Approach}, Springer, pages 11--22, 2019.

\bibitem[Ferrante et~al.(2021)Ferrante, Ferro, and
  Fuhr]{DBLP:journals/access/FerranteFF21}
Marco Ferrante, Nicola Ferro, and Norbert Fuhr.
\newblock Towards meaningful statements in {IR} evaluation: Mapping evaluation
  measures to interval scales.
\newblock \emph{{IEEE} Access}, 9:\penalty0 136182--136216, 2021.
\newblock \doi{10.1109/ACCESS.2021.3116857}.

\bibitem[Frihat and Fuhr(2025)]{frihat2025bioconcepts}
Sameh Frihat and Norbert Fuhr.
\newblock Integration of biomedical concepts for enhanced medical literature
  retrieval.
\newblock \emph{International Journal of Data Science and Analytics}, pages
  1--24, 2025.

\bibitem[Frihat et~al.(2024)Frihat, Papernmeier, and Fuhr]{JCDL:UserStudy2024}
Sameh Frihat, Papernmeier, and Norbert Fuhr.
\newblock Enhancing biomedical literature retrieval with level of evidence and
  bio-concepts: A comparative user study.
\newblock The ACM/IEEE Joint Conference on Digital Libraries (JCDL), 2024.

\bibitem[Fuhr(2017)]{Fuhr:17b}
Norbert Fuhr.
\newblock Some common mistakes in IR evaluation, and how they can be avoided.
\newblock \emph{SIGIR Forum},$\,$ 51\penalty0 (3):0 32--41,$\;$ 2017. $\;$
\newblock URL \url{http://sigir.org/wp-content/uploads/2018/01/p032.pdf}.

\bibitem[Group et~al.(2002)Group, Guyatt, Rennie, et~al.]{evidence2002users}
Evidence-Based Medicine~Working Group, Gordon Guyatt, Drummond Rennie, et~al.
\newblock \emph{Users' guides to the medical literature: a manual for
  evidence-based clinical practice}.
\newblock AMA Press, 2002.


\bibitem[Gu et~al.(2021)Gu, Tinn, Cheng, Lucas, Usuyama, Liu, Naumann, Gao, and
  Poon]{gu2021domain}
Yu~Gu, Robert Tinn, Hao Cheng, Michael Lucas, Naoto Usuyama, Xiaodong Liu,
  Tristan Naumann, Jianfeng Gao, and Hoifung Poon.
\newblock Domain-specific language model pretraining for biomedical natural
  language processing.
\newblock \emph{ACM Transactions on Computing for Healthcare (HEALTH)},
  3\penalty0 (1):\penalty0 1--23, 2021.

\bibitem[Guyatt(2009)]{guyatt2009grade}
GH~Guyatt.
\newblock Grade: an emerging consensus on rating quality of evidence and
  strength of recommendations.
\newblock \emph{Chinese Journal of Evidence-Based Medine}, 9:\penalty0 8, 2009.

\bibitem[Hartley(2004)]{hartley2004current}
James Hartley.
\newblock Current findings from research on structured abstracts.
\newblock \emph{Journal of the Medical Library Association}, 92\penalty0
  (3):\penalty0 368, 2004.

\bibitem[Hartling and Gates(2022)]{hartling2022friend}
Lisa Hartling and Allison Gates.
\newblock Friend or foe? the role of robots in systematic reviews.
\newblock \emph{Annals of Internal Medicine}, 175\penalty0 (7):\penalty0
  1045--1046, 2022.

\bibitem[Hirt et~al.(2021)Hirt, Meichlinger, Schumacher, and
  Mueller]{hirt2021agreement}
Julian Hirt, Jasmin Meichlinger, Petra Schumacher, and Gerhard Mueller.
\newblock Agreement in risk of bias assessment between robotreviewer and human
  reviewers: An evaluation study on randomised controlled trials in
  nursing-related cochrane reviews.
\newblock \emph{Journal of Nursing Scholarship}, 53\penalty0 (2):\penalty0
  246--254, 2021.

\bibitem[Howick(2011)]{howick2011oxford}
Jeremy Howick.
\newblock The Oxford 2011 levels of evidence.
\newblock \emph{Centre for Evidence-Based Medicine}, 2011.
\newblock $\,$ URL$\,$ \url{https://www.cebm.ox.ac.uk/resources/levels-of-evidence/ocebm-levels-of-evidence}

\bibitem[J{\"{a}}rvelin and
  Kek{\"{a}}l{\"{a}}inen(2000)]{DBLP:conf/sigir/JarvelinK00}
Kalervo J{\"{a}}rvelin and Jaana Kek{\"{a}}l{\"{a}}inen.
\newblock {IR} evaluation methods for retrieving highly relevant documents.
\newblock In Emmanuel~J. Yannakoudakis, Nicholas~J. Belkin, Peter Ingwersen,
  and Mun{-}Kew Leong, editors, \emph{{SIGIR} 2000: Proceedings of the 23rd
  Annual International {ACM} {SIGIR} Conference on Research and Development in
  Information Retrieval}, pages 41--48,
  {ACM}, 2000.
\newblock \doi{10.1145/345508.345545}.

\bibitem[Leung(2001)]{leung2001evidence}
Gabriel~M Leung.
\newblock Evidence-based practice revisited.
\newblock \emph{Asia Pacific Journal of Public Health}, 13\penalty0
  (2):\penalty0 116--121, 2001.

\bibitem[Marshall and Wallace(2019)]{marshall2019toward}
Iain~J Marshall and Byron~C Wallace.
\newblock Toward systematic review automation: a practical guide to using
  machine learning tools in research synthesis.
\newblock \emph{Systematic reviews}, 8:\penalty0 1--10, 2019.

\bibitem[Marshall et~al.(2014)Marshall, Kuiper, and
  Wallace]{marshall2014automating}
Iain~J Marshall, Jo{\"e}l Kuiper, and Byron~C Wallace.
\newblock Automating risk of bias assessment for clinical trials.
\newblock In \emph{Proceedings of the 5th ACM Conference on Bioinformatics,
  Computational Biology, and Health Informatics}, pages 88--95, 2014.

\bibitem[Marshall et~al.(2016)Marshall, Kuiper, and
  Wallace]{marshall2016robotreviewer}
Iain~J Marshall, Jo{\"e}l Kuiper, and Byron~C Wallace.
\newblock Robotreviewer: evaluation of a system for automatically assessing
  bias in clinical trials.
\newblock \emph{Journal of the American Medical Informatics Association},
  23\penalty0 (1):\penalty0 193--201, 2016.

\bibitem[Murad et~al.(2016)Murad, Asi, Alsawas, and Alahdab]{murad2016new}
M~Hassan Murad, Noor Asi, Mouaz Alsawas, and Fares Alahdab.
\newblock New evidence pyramid.
\newblock \emph{BMJ Evidence-Based Medicine}, 21\penalty0 (4):\penalty0
  125--127, 2016.

\bibitem[Pasche et~al.(2020)Pasche, Caucheteur, Mottin, Mottaz, Gobeill, and
  Ruch]{pasche2020sib}
Emilie Pasche, D{\'e}borah Caucheteur, Luc Mottin, Ana{\"\i}s Mottaz, Julien
  Gobeill, and Patrick Ruch.
\newblock SIB text mining at TREC precision medicine 2020.
\newblock In \emph{Proceedings of the 29th Text REtrieval Conference
  (TREC 2020)}. 16--20, 2020.

\bibitem[Patrick et~al.(2004)Patrick, Demiris, Folk, Moxley, Mitchell, and
  Tao]{patrick2004evidence}
Timothy~B Patrick, George Demiris, Lillian~C Folk, David~E Moxley, Joyce~A
  Mitchell, and Donghua Tao.
\newblock Evidence-based retrieval in evidence-based medicine.
\newblock \emph{Journal of the Medical Library Association}, 92\penalty0
  (2):\penalty0 196, 2004.

\bibitem[Polikar(2012)]{polikar2012ensemble}
Robi Polikar.
\newblock Ensemble learning.
\newblock \emph{Ensemble machine learning: Methods and applications}, pages
  1--34, 2012.

\bibitem[Ribeiro et~al.(2016)Ribeiro, Singh, and Guestrin]{lime2016why}
Marco~Tulio Ribeiro, Sameer Singh, and Carlos Guestrin.
\newblock ``Why should I trust you?'' explaining the predictions of any
  classifier.
\newblock In \emph{Proceedings of the 22nd ACM SIGKDD international conference
  on knowledge discovery and data mining}, pages 1135--1144, 2016.

\bibitem[Roberts et~al.(2017)Roberts, Demner-Fushman, Voorhees, Hersh, Bedrick,
  Lazar, and Pant]{roberts2017overview}
Kirk Roberts, Dina Demner-Fushman, Ellen~M Voorhees, William~R Hersh, Steven
  Bedrick, Alexander~J Lazar, and Shubham Pant.
\newblock Overview of the {TREC} 2017 precision medicine track.
\newblock In \emph{Proceedings of the 28th Text REtrieval Conference (TREC)}, 2017.

\bibitem[Roberts et~al.(2018)Roberts, Demner-Fushman, Voorhees, Hersh, Bedrick,
  and Lazar]{roberts2018overview}
Kirk Roberts, Dina Demner-Fushman, Ellen~M Voorhees, William~R Hersh, Steven
  Bedrick, and Alexander~J Lazar.
\newblock Overview of the {TREC} 2018 precision medicine track.
\newblock In \emph{Proceedings of the 29th Text REtrieval Conference (TREC)}, 2018.

\bibitem[Roberts et~al.(2019)Roberts, Demner-Fushman, Voorhees, Hersh, Bedrick,
  Lazar, Pant, and Meric-Bernstam]{roberts2019overview}
Kirk Roberts, Dina Demner-Fushman, Ellen~M Voorhees, William~R Hersh, Steven
  Bedrick, Alexander~J Lazar, Shubham Pant, and Funda Meric-Bernstam.
\newblock Overview of the {TREC} 2019 precision medicine track.
\newblock In \emph{Proceedings of the 30th Text REtrieval Conference (TREC)}, 2019.

\bibitem[Robertson et~al.(2009)Robertson, Zaragoza,
  et~al.]{robertson2009probabilistic}
Stephen Robertson, Hugo Zaragoza, et~al.
\newblock The probabilistic relevance framework: {BM25} and beyond.
\newblock \emph{Foundations and Trends in Information
  Retrieval}, 3\penalty0 (4):\penalty0 333--389, 2009.

\bibitem[Rosner(2012)]{rosner2012evidence}
Anthony~L Rosner.
\newblock Evidence-based medicine: revisiting the pyramid of priorities.
\newblock \emph{Journal of Bodywork and Movement Therapies}, 16\penalty0
  (1):\penalty0 42--49, 2012.

\bibitem[Sathian et~al.(2010)Sathian, Sreedharan, Baboo, Sharan, Abhilash, and
  Rajesh]{sathian2010relevance}
Brijesh Sathian, Jayadevan Sreedharan, Suresh~N Baboo, Krishna Sharan,
  ES~Abhilash, and E~Rajesh.
\newblock Relevance of sample size determination in medical research.
\newblock \emph{Nepal Journal of Epidemiology}, 1\penalty0 (1):\penalty0 4--10,
  2010.

\bibitem[Soboczenski et~al.(2019)Soboczenski, Trikalinos, Kuiper, Bias,
  Wallace, and Marshall]{soboczenski2019machine}
Frank Soboczenski, Thomas~A Trikalinos, Jo{\"e}l Kuiper, Randolph~G Bias,
  Byron~C Wallace, and Iain~J Marshall.
\newblock Machine learning to help researchers evaluate biases in clinical
  trials: a prospective, randomized user study.
\newblock \emph{BMC Medical Informatics and Decision Making}, 19:\penalty0
  1--12, 2019.

\bibitem[Strage et~al.(2023)Strage, Stacey, Mauffrey, and
  Parry]{strage2023interobserver}
Katya Strage, Stephen Stacey, Cyril Mauffrey, and Joshua~A Parry.
\newblock The interobserver reliability of clinical relevance in medical
  research.
\newblock \emph{Injury}, 54:\penalty0 S66--S68, 2023.

\bibitem[Van~de Vliet et~al.(2023)Van~de Vliet, Sprenger, Kampers, Makalowski,
  Schirrmacher, St{\"u}cker, and Van~Gool]{van2023application}
Peter Van~de Vliet, Tobias Sprenger, Linde~FC Kampers, Jennifer Makalowski,
  Volker Schirrmacher, Wilfried St{\"u}cker, and Stefaan~W Van~Gool.
\newblock The application of evidence-based medicine in individualized
  medicine.
\newblock \emph{Biomedicines}, 11\penalty0 (7):\penalty0 1793, 2023.

\bibitem[Zhou and Zhou(2021)]{zhou2021ensemble}
Zhi-Hua Zhou and Zhi-Hua Zhou.
\newblock \emph{Ensemble learning}.
\newblock Springer, 2021.

\end{thebibliography}

\end{document}